\documentclass[10pt,letterpaper,twocolumn]{article} %% two column, final layout

\usepackage{ol2}
\usepackage[draft]{hyperref}
\usepackage{amsmath}

\begin{document}

\twocolumn[ %% activate for two-column option

\title{Half-spectral unidirectional invisibility in non-Hermitian periodic optical structures}

%% For REVTeX it is possible to automate superscript and e-mail callouts with the superscriptaddress option; see REVTeX4 documentation.

\author{Stefano Longhi}

\address{Dipartimento di Fisica, Politecnico di Milano and Istituto di Fotonica e Nanotecnologie del Consiglio Nazionale delle Ricerche, Piazza L. da Vinci 32, I-20133 Milano, Italy (stefano.longhi@polimi.it)}

\begin{abstract}
The phenomenon of {\it half-spectral} unidirectional invisibility is introduced for one-dimensional periodic optical structures with tailored real and imaginary refractive index distributions in a non-$\mathcal{PT}$-symmetric configuration. The effect refers to the property that the optical medium appears to be invisible, both in reflection and transmission, below the Bragg frequency when probed from one side, and above the Bragg frequency when probed from the opposite side. Half-spectral invisibility is obtained by a combination of in-phase index and gain gratings whose spatial amplitudes are related each other by a Hilbert transform.
\end{abstract}


 ] %% activate for two-column option
  
In recent years considerable research efforts have
been devoted to the design and realization of synthetic optical media that exploit the domain of {\it complex} dielectric permittivity $\epsilon$ 
to mold the flow of light in rather unusual ways (see, for example, \cite{r1,r2,r3,r4,r5,r6,r6bis,r7,r8,r9,r10,r11,r12,r13,r13b} and references therein). In such media, the presence of gain and loss makes wave dynamics non-Hermitian in nature. When special tailoring of the spatial distribution of  the real and imaginary parts of $\epsilon$ is introduced, important phenomena and devices can be realized, such as double refraction \cite{r1}, unidirectional and anomalous transport \cite{r2,r6,r7,r13b}, laser-absorber structures \cite{r4,r5,r9},
bidirectional \cite{r14} and
 unidirectional \cite{r15,r16,r17,r18,r19,r20} transparency,  laser mode selection \cite{r10,r11},  invisible defects \cite{r21,r22}, and reflectionless interfaces \cite{r13}, to mention a few. Often a balance spatial distribution of loss and gain is introduced to cancel each other on  average in so-called parity-time ($\mathcal{PT}$) symmetry configuration inspired by non-Hermitian quantum mechanics \cite{r1,r23,r24}, however this is not a necessary requirement \cite{r13,r21}.\par
An important class of non-Hermitian optical media is provided by one-dimensional
structures with locally periodic and complex $\epsilon$. In optics, periodic structures are the building blocks of many important optical devices, including
Bragg mirrors, optical filters, and distributed-feedback
lasers. In complex gratings, Bragg scattering is strongly influenced by the interplay between the real and imaginary parts of the locally-periodic refractive index distribution \cite{r2,r14,r15,r16,r17,r18,r19,r20,r25}. In particular, it has been shown that in $\mathcal{PT}$-symmetric gratings unidirectional invisibility can be observed near the symmetry breaking point.\par
In this Letter we introduce the phenomenon of {\it half-spectral} unidirectional invisibility in Bragg gratings with combined index and gain gratings tailored in a non-$\mathcal{PT}$-symmetric configuration. The effect refers to the property that the grating turns out to be invisible when probed from one side with an optical wave  at carrier frequency $\omega$ below the Bragg frequency $\omega_B$,  i.e. for $\omega< \omega_B$, whereas for $\omega>\omega_B$ the grating is invisible when probed from the opposite side.\par
Let us consider wave propagation in a one-dimensional
periodic optical structure with a complex refractive index profile
$n(X)=n_0+\Delta n h(X) \cos(2 \pi X / \Lambda)$, where $\Lambda$ is
the spatial period of the grating, $n_0$ is the (real-valued)
refractive index in absence of the grating, $\Delta n \ll n_0 $ is the  
the grating depth, and $h(X) \sim 1 $ is the amplitude profile of the grating.
The envelope $h(X)$ is assumed to be complex-valued, so that the real and imaginary parts of $h(X)$ correspond to 
the profiles of the index and gain/loss gratings inscribed in the medium. Note that the grating structure is not locally $\mathcal{PT}$ symmetric, since the index and gain/loss gratings are in phase rather than $\pi/2$ phase shifted, however gain and loss regions are balanced. The periodic
modulation of the refractive index leads to
Bragg scattering between two counterpropagating waves at frequency $\omega$ 
close to the Bragg frequency $\omega_B=\pi c_0/(\Lambda n_0)$, where
$c_0$ is the speed of light in vacuum [Fig.1(a)]. Indicating by
$E(X,t)=\psi_1(X,t) \exp(ik_BX-i \omega_B t)+\psi_2(X,t)\exp(-ik_BX-i
\omega_Bt)+c.c.$ the electric field propagating in the grating,
where $k_B=\pi/ \Lambda$, the slowly-varying envelopes $\psi_1$  and
$\psi_2$ of counter-propagating waves satisfy the coupled-mode
equations \cite{r26}. After introduction of dimensionless spatial and temporal variables $x=X / \mathcal{L}$
and $\tau=t/ \mathcal{T}$ with characteristic spatial and time scales $\mathcal{L} \equiv 2 n_0 \Lambda /(\pi \Delta n)$ and 
$\mathcal{T}=\mathcal{L}n_0/c_0$, coupled mode equations can be cast in the Dirac-type form for the spinor $\psi \equiv (\psi_1,\psi_2)^T$
 \cite{r27}
\begin{equation}
i \frac{\partial \psi}{\partial \tau}= i \left(
\begin{array}{cc}
-1 & 0 \\
0 & 1
\end{array}
\right)
\frac{\partial \psi}{\partial x}-\left(
\begin{array}{cc}
0 & h(x) \\
h(x) & 0
\end{array}
\right) \psi,
\end{equation}
where $h(x)$ plays the role of a complex variable mass (scalar coupling). We assume that $h(x) \rightarrow 0$ as $|x| \rightarrow \infty$ sufficiently fast, so that one can introduce the transmission $t(\delta)$ and reflection $r^{(l,r)}(\delta)$ coefficients (for left and right incidence sides), where $\delta=(2 n_0 / \Delta n)(\omega/ \omega_B-1)$ is the normalized frequency detuning of incident wave from the Bragg frequency $\omega_B$. In the Dirac equation (1), the condition $h(x) \rightarrow 0$ as $|x| \rightarrow \infty$ means that the particle is asymptotically massless, and the cases $\delta>0$ and $\delta<0$ correspond to positive/negative energy states, i.e. to particle/anti-particle states, respectively. To determine the transmission and reflection coefficients, let us set $\psi_1(x,\tau)=u(x) \exp(-i \delta \tau+i \delta x)$ and $\psi_2(x,\tau)=v(x) \exp(-i \delta \tau-i \delta x)$, so that the amplitudes $u,v$ satisfy the coupled equations
\begin{eqnarray}
\frac{du}{dx} & = & ih(x) \exp(-2i \delta x) v \\
\frac{dv}{dx} & = & -ih(x) \exp(2i \delta x) u.
\end{eqnarray}
\begin{figure}[htb]
\centerline{\includegraphics[width=8.4cm]{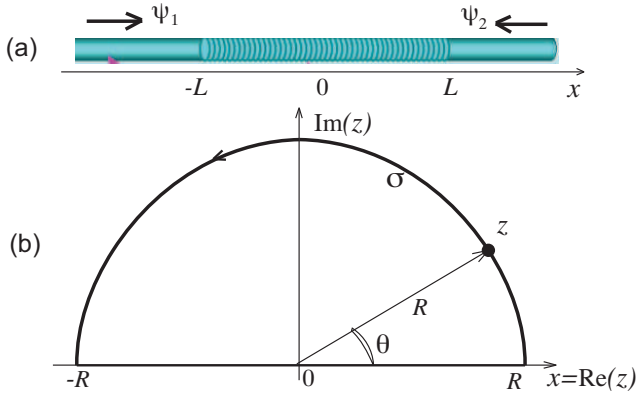}} \caption{ \small
(Color online) (a) Schematic of Bragg scattering of counter-propagating waves in a mixed index/gain grating.
(b) Contour path $\sigma$ in the complex $z$ plane  (parametric equation $z=R \exp(i \theta)$, $0 \leq \theta \leq \pi$) used for the integration of Eqs.(2) and (3).}
\end{figure}
 \begin{figure}[htb]
\centerline{\includegraphics[width=8.6cm]{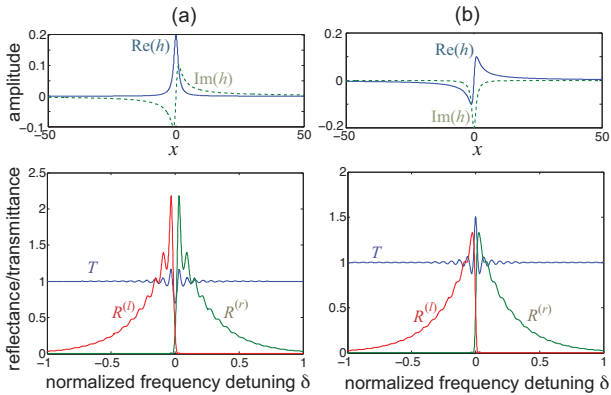}} \caption{ \small
(Color online) Numerically-computed behavior of spectral transmittance $T$ and reflectances $R^{(l,r)}$ for left and right incidence sides in a periodic grating with complex envelope $h(x)$ defined by Eq.(13) for (a) $A=0.2i$, $\alpha=1$, and (b) $A=0.2$, $\alpha=1$. The grating length is $2L=100$. The upper panels in (a) and (b) show the behavior of the real and imaginary parts of $h(z)$, corresponding to the envelopes of index and gain gratings, respectively.}
\end{figure}
 \begin{figure}[htb]
\centerline{\includegraphics[width=8.6cm]{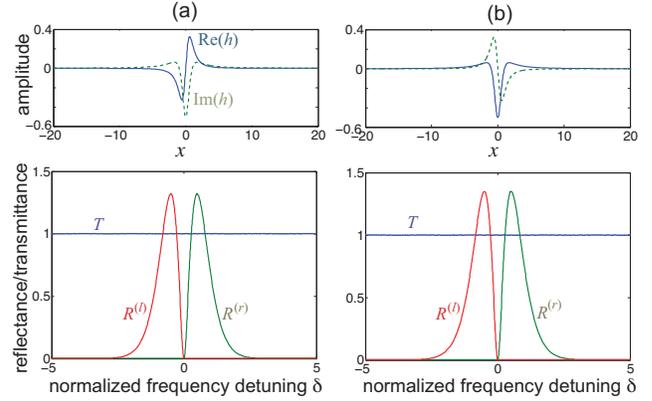}} \caption{ \small
(Color online)  (Color online) Same as Fig.2, but for a grating with complex envelope $h(x)$ defined by Eq.(14). (a) $A=0.5i$, $\alpha=1$, and (b) $A=0.5$, $\alpha=1$. The grating length is $2L=100$.}
\end{figure}
The transmission and reflection coefficients are formally defined as 
$t(\delta)=u(+\infty)$ and $r^{(l)}(\delta)=v(-\infty)$ with the boundary conditions $u(-\infty)=1$, $v(+\infty)=0$ for left side incidence, and 
$t(\delta)=v(x=-\infty)$ and $r^{(r)}(\delta)=u(+\infty)$ with the boundary conditions $u(-\infty)=0$, $v(+\infty)=1$ for right side incidence. Note that the transmission coefficient does not depend on the incidence side, while the reflection coefficient does. We say that the grating shows half-spectral unidirectional invisibility whenever $t(\delta)=1$ (for $\delta \neq 0$) and $r^{(l)}(\delta)=0$ for $\delta>0$, $r^{(r)}(\delta)=0$ for $\delta<0$. Interestingly, within the Dirac formulation of the scattering problem \cite{r27} this means that the scalar coupling $h(x)$ is invisible for forward-propagating positive energy states (particles), whereas it is invisible for backward-propagating negative energy states (anti-particles). The main result of the present Letter is that, whenever the real and imaginary parts of the complex grating envelope $h(x)$ are related by a Hilbert transform, i.e. they satisfy spatial Kramers-Kronig relations \cite{r13}, half-spectral unidirectional invisibility is observed. Since the real and imaginary parts of $h(x)$ are related by  a Hilbert transform, $h(x)$ can be analytically prolonged into the upper half complex, whereas $h$ has some poles in the lower half complex plane.  
To prove the half-spectral invisibility property, let us extend Eqs.(2,3) into the complex $z$ plane, and let us assume the asymptotic behavior $h(z) \sim M/ z^ \gamma$ for some $M>0$ and $\gamma \geq 1$ as $|z| \rightarrow \infty$, where $\gamma$ is the degree of the lowest-order pole of $h(z)$ in the ${\rm Im}(z)<0$ complex plane.
 Let us indicate by $u_0(z)$, $v_0(z)$ the solution to Eqs.(2,3) with the asymptotic behavior $u_0(z) \rightarrow 1$ and $v_0(z) \rightarrow 0$ as $z=x$ real $\rightarrow + \infty$. Since the real and imaginary parts of $h(z)$ are related by a Hilbert transform, $h(z)$ is holomorphic in the ${\rm Im}(z) \geq 0$ half plane, and thus the functions  $u_0(z)$  and $ v_0(z)$ are holomorphic in the same domain as well. To calculate the asymptotic behavior of the solution $u_0(z)$, $ v_0(z)$ as $z=x$ real $\rightarrow - \infty$, we can integrate Eqs.(2,3) along the semi-circumference $\sigma$ of radius $R \rightarrow \infty$, described by the equation $z=R \exp(i \theta)$ with $0 \leq \theta \leq \pi$ [Fig.1(b)]. Since the solution is analytic in the entire ${\rm Im} (z) \geq 0$ domain and there are no branch cuts, the integration of Eqs.(2,3) along the path $\sigma$ and along the real axis $z=x$ yields the same result. After setting $z=R \exp(i \theta)$ in Eqs.(2,3), one obtains
\begin{eqnarray}
\frac{du_0}{d \theta} & = & \kappa_{1,2} (\theta) v_0 \\
\frac{dv_0}{d \theta} & = & \kappa_{2,1} (\theta) u_0
\end{eqnarray}
 which should be integrated from $\theta=0$ to $\theta= \pi$ with the initial condition $u_0(\theta=0)=1$, $v_0(\theta=0)=0$. In Eqs.(4,5), the couplings $\kappa$ are defined by
 \begin{eqnarray}
 \kappa_{1,2} (\theta)  =  -R h \left(  R \exp(i \theta) \right) \exp \left[ i \theta-2i \delta R \exp(i \theta) \right]  \\
 \kappa_{2,1} (\theta)  =  R h \left(  R \exp(i \theta) \right) \exp \left[ i \theta+2i \delta R \exp(i \theta) \right].
\end{eqnarray}
Note that, for $\delta >0$ and for a sufficiently large radius $R$ one has
$|\kappa_{2,1} (\theta)| < M R^{1-\gamma} \exp(-2 \delta R \sin \theta) \rightarrow 0$, wehereas
$\kappa_{1,2} \kappa_{2,1}$ is limited or vanishes as $R \rightarrow \infty$. To determine the solution to Eqs.(4,5) is the asymptotic limit $R \rightarrow \infty$, it is worth eliminating the variable $v_0$, obtaining the second-order equation for $u_0(\theta)$
\begin{eqnarray}
\frac{d^2u_0}{d \theta^2}+ \left[  i(\gamma-1) -2 \delta R \exp( i \theta) \right] \frac{du_0}{d \theta} + \nonumber \\
+\frac{M^2}{R^{2(\gamma-1)}} \exp [2i(1-\gamma) \theta] u_0=0
\end{eqnarray}
which should be integrated with the initial condition $u_0(0)=1$, $(d u_0 / d \theta)(0)=0$. In the large limit $R \rightarrow \infty$ and for $ \delta \neq 0$, Eq.(8) can be approximated as
$(d^2 u_0 / d \theta^2)-2 \delta R \exp(i \theta) (d u_0 /d \theta) \simeq 0$. The solution to this equation with the given initial condition is simply given by $u_0(\theta)=1$. Moreover, from Eq.(5) one has $v_0(\theta)=\int_0^ \theta  \kappa_{2,1}( \xi) d \xi \rightarrow 0$ when $\delta>0$ and $R \rightarrow \infty$. Hence, along the semi-circumference $\sigma$ with $R \rightarrow \infty$ radius and $\delta >0$, one has $v_0( \theta)=0$ and $u_0(\theta)=1$, and thus $\lim_{x \rightarrow -\infty}u_0(x)=1$ and $\lim_{x \rightarrow -\infty} v_0(x)=0$ on the real $x$ axis. This means that $t(\delta)=1$ and $r^{(l)}(\delta)=0$ for $\delta>0$. In a similar way one can show that $t(\delta)=1$ and $r^{(r)}(\delta)=0$ for $\delta<0$. The case $\delta=0$ (exact Bragg resonance $\omega= \omega_B$) is somehow a critical one and should be considered separately. In this case, to calculate the transmission and reflection coefficients let us truncate the grating at the distances $x=\pm L$ [see Fig.1(a)], taking at the end of the calculations the limit $L \rightarrow \infty$. For $\delta=0$ Eqs.(2,3) can be exactly integrated from $x=-L$ to $x=L$, yielding the exact input-output relation $(u,v)^T_{x=L}= \mathcal{M} (u,v)^T_{x=-L}$ with the transfer matrix $\mathcal{M}$ given by
\begin{equation}
\mathcal{M}= \left(
\begin{array}{cc}
\cosh \left(\int_{-L}^{L} dx h(x) \right) & i \sinh \left(\int_{-L}^{L} dx h(x) \right) \\
-i \sinh \left(\int_{-L}^{L} dx h(x) \right) & \cosh \left(\int_{-L}^{L} dx h(x) \right)
\end{array}
\right).
\end{equation}
 \begin{figure}[htb]
\centerline{\includegraphics[width=8.6cm]{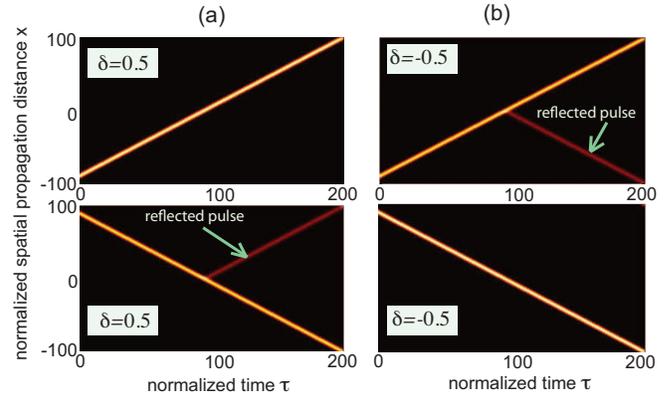}} \caption{ \small
(Color online) Propagation of a Gaussian pulse (snapshot of the intensity $|\psi_1|^2+|\psi_2|^2$ in the space-time $(x,\tau)$ plane) across the optical grating of Fig.2(a) for (a) $\delta=0.5$, and (b) $\delta=-0.5$. Upper and lower panels refer to left and right sides of incidence, respectively.}
\end{figure}
\begin{figure}[htb]
\centerline{\includegraphics[width=8.6cm]{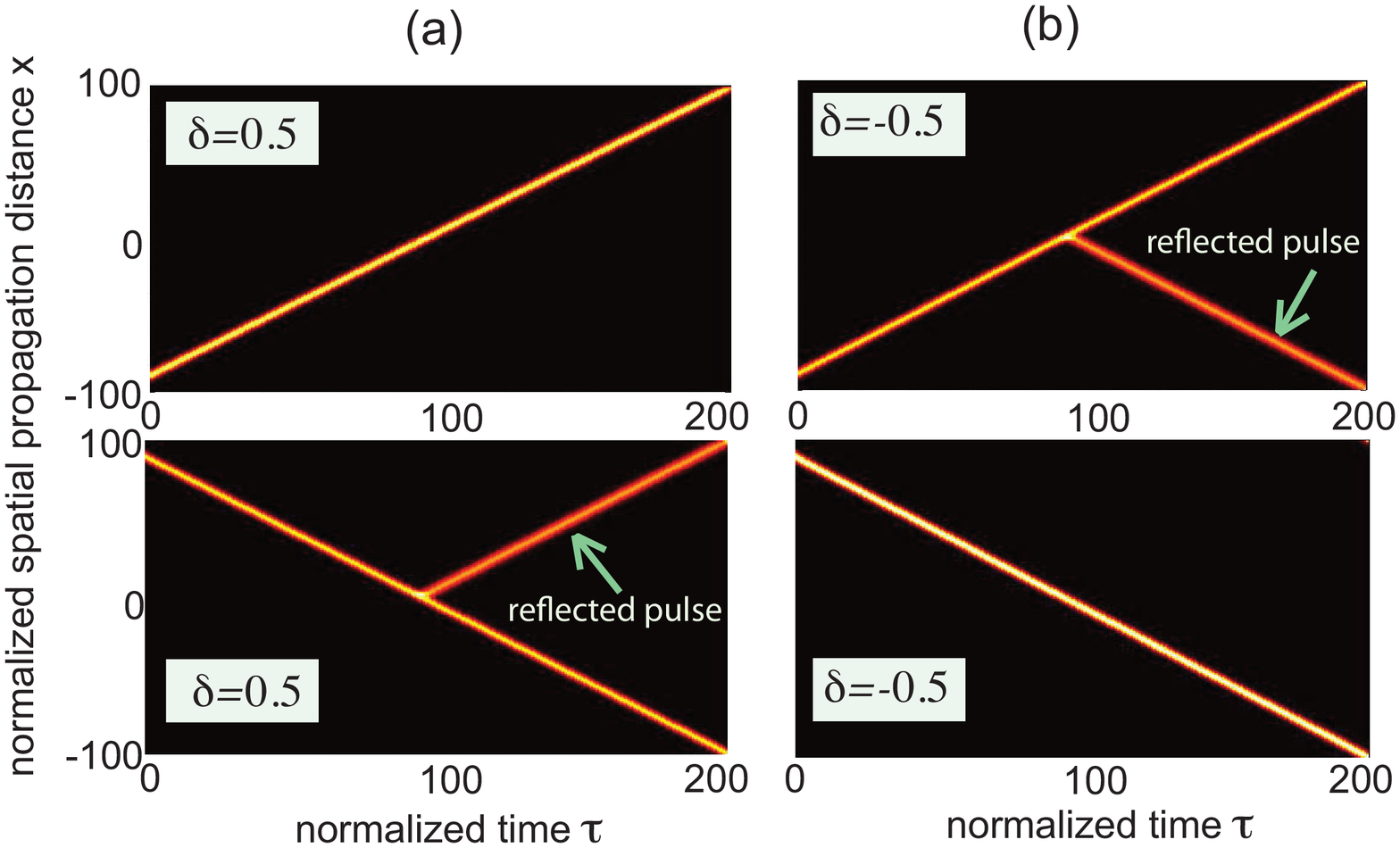}} \caption{ \small
(Color online) Propagation of a Gaussian pulse (snapshot of the intensity $|\psi_1|^2+|\psi_2|^2$ in the space-time plane) across the optical grating of Fig.3(a) for (a) $\delta=0.5$, and (b) $\delta=-0.5$. Upper and lower panels refer to left and right sides of incidence, respectively.}
\end{figure}
The transmission and reflection coefficients are expressed in terms of the elements of the transfer matrix by usual relations, yielding
\begin{equation}
t(\delta=0)=\frac{1}{\mathcal{M}_{22}}=\frac{1}{\cosh \left(\int_{-L}^{L} dx h(x) \right)}
\end{equation}
\begin{equation}
r^{(l)}(\delta=0)=-\frac{\mathcal{M}_{21}
}{\mathcal{M}_{22}}=i \tanh \left(\int_{-L}^{L} dx h(x) \right)
\end{equation}
\begin{equation}
r^{(r)}(\delta=0)=\frac{\mathcal{M}_{12}
}{\mathcal{M}_{22}}=r^{(l)}(\delta=0).
\end{equation}
The values of the transmission/reflection coefficients at Bragg resonance are thus determined  by the limit $\lim_{L \rightarrow \infty} \int_{-L}^{L} h(x) dx$, which is given by the sum of residues of poles of $h(z)$ in the lower half complex plane, multiplied by $ -i \pi$. If the sum of residues is zero or whenever the poles of $h(z)$ are of order higher than one, one has $t(\delta=0)=1$, $r^{(l,r)}(\delta=0)=0$ and the function $t(\delta)$, $r^{(l,f)}(\delta)$ are continuous ones when crossing the resonance condition $\delta=0$. Conversely, when the integral $\int_{-\infty}^{\infty} h(x) dx$ is non-vanishing one, a discontinuity in the transmittance and reflectances at $\delta=0$ arises. As an example, Figs.2 and 3 show the numerically-computed behavior of the spectral transmittance $T=|t|^2$ and reflectances $R^{(l,r)}=|r^{(l,r)}|^2$ versus $\delta$ for the two complex profiles 
\begin{eqnarray}
h(x) & = & \frac{A}{x+i \alpha} \\
h(x) & = & \frac{A}{(x+i \alpha)^2} 
\end{eqnarray}
($\alpha>0$), corresponding to a single pole of first and second order, respectively. The profiles  of the real and imaginary parts of $h(x)$, defining the envelopes of the index and gain/loss gratings, are also shown in the figure. Note that for the case of a first-order pole (Fig.2), the transmittance rapidly oscillates close to the resonance $\delta=0$, settling down to $t=1$ outside a narrow range around $\delta=0$. Such a result is due to grating truncation and to the non-vanishing value of $\int_{-L}^L h(x) dx$, corresponding to $t(\delta=0) \neq 1$ according to Eq.(10). Note also that the reflectances for left and ride incidence sides satisfy the symmetry relation $R^{(l)}(-\delta)=R^{(r)}(\delta)$. Quite interestingly,  especially in the case of a pole of first-order (Fig.2) the reflectance shows a steep increase from zero to a non-negligible value as the detuning $\delta$ is swept across the resonance $\delta=0$. Such a steep change of reflectance could be of usefulness, for example, in sensing applications, where perturbation-induced  small shift of the resonance condition leads to a large change in the grating reflectivity at a given probing  wavelength.\\ 
It should be noted that, since the optical medium displays a gain grating superimposed to an index grating, unstable bound states, corresponding to lasing modes, could be sustained by the optical structure despite gain and loss are balanced on average \cite{r14}. For the observation of half-spectral unidirectional invisibility, the appearance of unstable lasing modes  should be avoided. We numerically checked the absence of instabilities by direct numerical simulations of pulse propagation  in the grating structures shown in Figs.2 and 3 based on coupled-mode equations (1). Figures 4 and 5 show the 
temporal evolution of  $|\psi_1(x,\tau)|^2+| \psi_2(x,\tau)|^2$ - which
is proportional to the field intensity in the grating averaged in time over a
few optical cycles and in space over a few wavelengths - as
obtained by numerical analysis of Eq.(1) assuming as an initial condition either a forward-propagating or a backward-propagating
Gaussian pulse of duration (FWHM in intensity)
$\tau_p=5$ and with frequency detuning $\delta=0.5$ [in panels (a)] and $\delta=-0.5$ [in panels (b)]. The figures clearly demonstrate the observation of the phenomenon of half-spectral unidirectional invisibility. It should be noted that, for a fixed value of the pole position $z=-i \alpha$ and grating length $2L$, as the amplitude $|A|$ in the profile given by Eq.(13) (first-order pole) is increased, unstable modes are observed, whereas no instabilities are found in the second-order pole profile Eq.(14). The onset of an instability is signaled  by the appearance of singularities (poles) in the spectral transmittance as $|A|$ is increased \cite{r27}. Figure 6 shows the numerically-computed spectral transmittance for the grating profile defined by Eq.(13) for $\alpha=1$ and for increasing values of $|A|$ [real values of $A$ in panel (a), imaginary values of $A$ in panel (b)]. Note that, for real $A$ a singularity sets in at $\delta=0$ and $A \sim 0.51$, whereas for imaginary $A$ two singularities are found at $\delta= \pm 0.0354$ and $A \sim i 0.62$.

\begin{figure}[htb]
\centerline{\includegraphics[width=8.6cm]{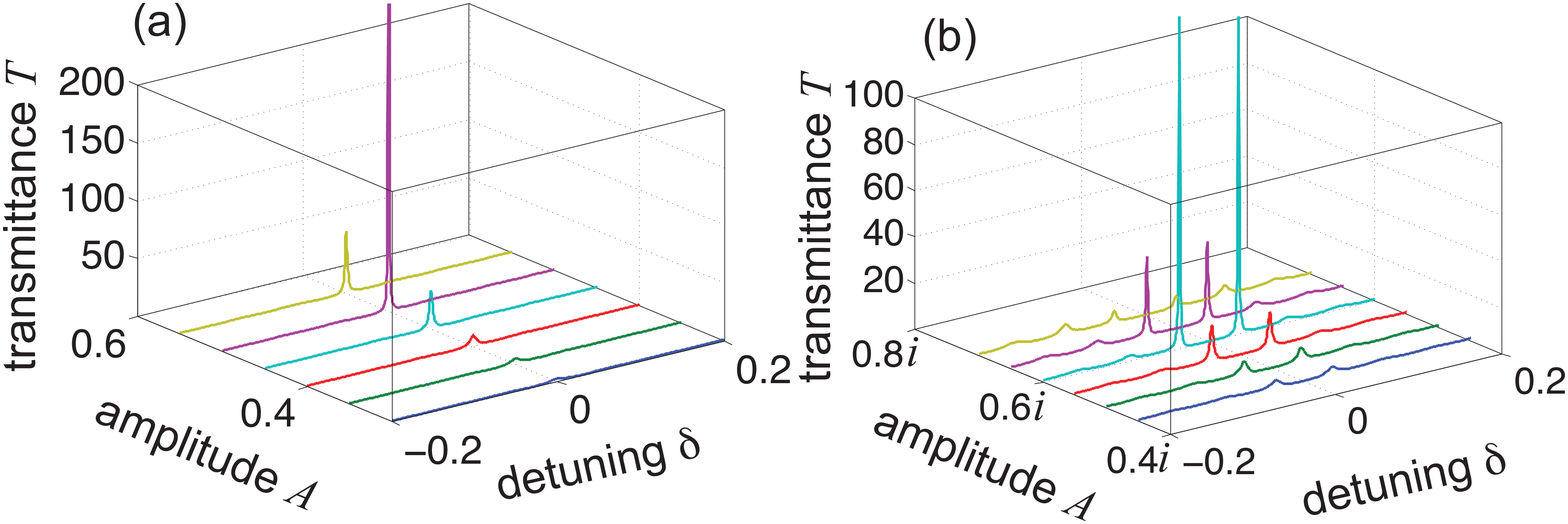}} \caption{ \small
(Color online) Numerically-computed behavior of spectral transmittance $T$ for the grating profile defined by Eq.(13) for $\alpha=1$ and for a few increasing values of the amplitude $A$. In (a) $A$ is real, in (b) $A$ is imaginary.  The grating length is $2L=100$. Divergences in the spectral transmittance signal the onset of instabilities.}
\end{figure}

 In conclusion, the phenomenon of half-spectral unidirectional invisibility has been introduced for one-dimensional periodic optical structures with tailored index and gain gratings in a non-$\mathcal{PT}$-symmetric configuration. We have shown that, whenever the index and gain grating profiles are in phase and related each other by a Hilbert transform, i.e. by spatial Kramers-Kronig relations \cite{r13}, the optical medium appears  fully invisible from one side when probed below the Bragg frequency, and fully invisible from the other side when probed above the Bragg frequency. Our results disclose an important property of periodic optical structures with mixed index and gain gratings related by a Hilbert transform. Besides to provide a new  effect in the rapidly-growing field of non-Hermitian and $\mathcal{PT}$-symmetric optics, half-spectral transparency found in such structures might be of interest in applications, for the example in the design of sharp half-band optical filters. Moreover, the steep change of the spectral reflectance close to the Bragg frequency could be of usefulness in sensing applications.

\newpage

%%%%%%%%%%%%%%%%%%%%%%%%%%%%%%%
% References with full titles %
%%%%%%%%%%%%%%%%%%%%%%%%%%%%%%%

%\footnotesize
 {\bf References with full titles}\\
 \\
1. K. G. Makris, R. El-Ganainy, D. N. Christodoulides, and Z. H. Musslimani, {\it Beam Dynamics in $\mathcal{PT}$-Symmetric Optical Lattices}, Phys. Rev. Lett. {\bf 100}, 103904 (2008).\\
2. S. Longhi, {\it Bloch Oscillations in Complex Crystals with $\mathcal{PT}$ Symmetry}, Phys. Rev. Lett. {\bf 103}, 123601 (2009).\\ 
3. C. E. R\"{u}ter, K.G. Makris, R. El-Ganainy, D.N. Christodoulides, M. Segev, and D. Kip, {\it Observation of parity-time symmetry in optics}, Nature Phys. {\bf 6}, 192 (2010).\\
4. S. Longhi, {\it $\mathcal{PT}$ symmetric laser-absorber}, Phys. Rev. A {\bf 82}, 031801 (2010).\\
5. Y.D. Chong, L. Ge, and A.D. Stone, {\it $\mathcal{PT}$-symmetry breaking and laser-absorber modes in optical scattering systems}, Phys. Rev. Lett. {\bf 106}, 093902 (2011).\\
6. A. Regensburger, C. Bersch, M.-A. Miri, G. Onishchukov, D.N. Christodoulides, and U. Peschel, {\it Parity-time synthetic photonic lattices}, Nature {\bf 488}, 167 (2012).\\
7.  A. A. Sukhorukov, S. V. Dmitriev, S. V. Suchkov, and Yu. S. Kivshar, {\it Nonlocality in $\mathcal{PT}$-symmetric waveguide arrays with gain and loss}, Opt. Lett. {\bf 37}, 2148 (2012).\\
8. T. Eichelkraut, R. Heilmann, S. Weimann, S. St\"{u}tzer, F. Dreisow, D.N. Christodoulides, S. Nolte, and A. Szameit, {\it Mobility transition from ballistic to diffusive transport in non-Hermitian lattices},  Nature Commun. {\bf 4}, 2533 (2013).\\
9. B. Peng, S. K. Ozdemir,	F. Lei, F. Monifi,	M. Gianfreda, G.L. Long, S. Fan, F. Nori, C.M. Bender, and L. Yang, 
 {\it Parity-time-symmetric whispering-gallery microcavities}, Nature Phys. {\bf 10}, 394 (2014).\\
10. S. Longhi and L. Feng, {\it $\mathcal{PT}$-symmetric microring laser absorber}, Opt. Lett. {\bf 39}, 5026 (2014).\\ 
11.  H. Hodaei, M.-A. Miri, M. Heinrich, D.N. Christodoulides, and M. Khajavikhan, {\it Parity-time-symmetric microring lasers}, Science {\bf 346},  975 (2014).\\
12. L. Feng, Z.J. Wong, R.-M. Ma, Y. Wang, and X. Zhang, {\it Single-mode laser by parity-time symmetry breaking}, Science {\bf 346}, 972 (2014).\\
13. M. Wimmer,	A. Regensburger,	M.-A. Miri,	C. Bersch, D.N. Christodoulides, and U. Peschel, {\it  Observation of optical solitons in $\mathcal{PT}$-symmetric lattices}, Nature Commun. {\bf 6}, 7782 (2015).\\
14. S.A.R. Horsley, M. Artoni, and G.C. La Rocca, {\it Spatial Kramers-Kronig relations and the reflection of waves}, Nature Photon. {\bf 9}, 436 (2015).\\
15. S. Longhi, D. Gatti, and G. Della Valle, {\it Robust light transport in non-Hermitian photonic lattices}, Sci. Rep. {\bf 5}, 13376  (2015).\\
16. S. Longhi, {\it Transparency in Bragg scattering and phase conjugation}, Opt. Lett. {\bf 35}, 3844 (2010).\\
17. M. Kulishov, J.M. Laniel, N. Belanger, J. Azana, and D.V. Plant, {\it Nonreciprocal waveguide Bragg gratings }, Opt. Express {\bf 13}, 3068 (2005).\\
18. Z. Lin, H. Ramezani, T. Eichelkraut, T. Kottos, H. Cao, and D.N. Christodoulides, {\it Unidirectional Invisibility Induced by $\mathcal{PT}$-Symmetric Periodic Structures}, Phys. Rev. Lett. {\bf 106}, 213901 (2011). \\
19. S. Longhi, {\it Invisibility in $\mathcal{PT}$-symmetric complex crystals}, J. Phys. A {\bf 44}, 485302 (2011).\\
20. H.F. Jones, {\it Analytic results for a $\mathcal{PT}$-symmetric optical structure}, J. Phys. A {\bf 45}, 1751 (2012).\\
21. L. Feng, Y.-L. Xu, W.S. Fegadolli, M.-H. Lu, J.E.B. Oliveira, V.R. Almeida, Y.-F. Chen, and A. Scherer, {\it Experimental demonstration of a unidirectional reflectionless parity-time metamaterial at optical frequencies},     Nature Mat. {\bf 12}, 108 (2013).\\
22. M. Kulishov, H. F. Jones, and B. Kress, {\it Analysis of $\mathcal{PT}$-symmetric volume gratings beyond the paraxial approximation}, Opt. Express {\bf 23}, 9347 (2015).\\
23.  A. Mostafazadeh, {\it Transfer matrices as nonunitary S matrices, multimode unidirectional invisibility, and perturbative inverse scattering} Phys. Rev. A {\bf 89}, 012709 (2014).\\
24. S. Longhi, {\it Invisibility in non-Hermitian tight-binding lattices}, Phys. Rev. A {\bf 82}, 032111 (2010).\\
25. S. Longhi and G. Della Valle, {\it Invisible defects in complex crystals}, Ann. Phys. {\bf 334}, 35 (2013).\\
26. A. Ruschhaupt, F. Delgado, and J.G. Muga, {\it Physical realization of $\mathcal{PT}$-symmetric potential scattering in a planar slab waveguide}, J. Phys. A  {\bf 38}, L171 (2005).\\
27. R. El-Ganainy, K. G. Makris, D. N. Christodoulides, and Z.H. Musslimani, {\it Theory of coupled optical $\mathcal{PT}$ symmetric structures}, Opt. Lett. {\bf 32}, 2632 (2007).\\
28. S. Longhi, {\it Spectral singularities and Bragg scattering in complex crystals}, Phys. Rev. A {\bf 81}, 022102 (2010).\\
29. L. Poladian, {\it Resonance mode expansions and exact solutions for nonuniform gratings}, Phys. Rev. E {\bf 54}, 2963 (1996).\\
30.  S. Longhi, {\it Optical Realization of Relativistic Non-Hermitian Quantum Mechanics}, Phys. Rev. Lett. {\bf 105}, 013903 (2010).\\
 

\begin{thebibliography}{99}


%%%%%%%%%%%%%%%%%%%%%%%%%%%%%%%
% References (short version)  %
%%%%%%%%%%%%%%%%%%%%%%%%%%%%%%%

\bibitem{r1}
 K. G. Makris, R. El-Ganainy, D. N. Christodoulides, and Z. H. Musslimani, Phys. Rev. Lett. {\bf 100}, 103904 (2008).
 \bibitem{r2}
S. Longhi, Phys. Rev. Lett. {\bf 103}, 123601 (2009).
\bibitem{r3}
C. E. R\"{u}ter, K.G. Makris, R. El-Ganainy, D.N. Christodoulides, M. Segev, and D. Kip, Nature Phys. {\bf 6}, 192 (2010).
\bibitem{r4}
 S. Longhi, Phys. Rev. A {\bf 82}, 031801 (2010).
 \bibitem{r5}
Y.D. Chong, L. Ge, and A.D. Stone, Phys. Rev. Lett. {\bf 106}, 093902 (2011).
\bibitem{r6}
A. Regensburger, C. Bersch, M.-A. Miri, G. Onishchukov, D.N. Christodoulides, and U. Peschel, Nature {\bf 488}, 167 (2012).
\bibitem{r6bis}
 A. A. Sukhorukov, S. V. Dmitriev, S. V. Suchkov, and Yu. S. Kivshar,
Opt. Lett. {\bf 37}, 2148 (2012).
\bibitem{r7}
 T. Eichelkraut, R. Heilmann, S. Weimann, S. St\"{u}tzer, F. Dreisow, D.N. Christodoulides, S. Nolte, and A. Szameit, Nature Commun. {\bf 4}, 2533 (2013).
 \bibitem{r8}
 B. Peng,	S. K. Ozdemir,	F. Lei,	F. Monifi,	M. Gianfreda,	G.L. Long, S. Fan, F. Nori,	C.M. Bender, and L. Yang, Nature Phys. {\bf 10}, 394 (2014).
 \bibitem{r9}
S. Longhi and L. Feng, Opt. Lett. {\bf 39}, 5026 (2014).
\bibitem{r10}
H. Hodaei, M.-A. Miri, M. Heinrich, D.N. Christodoulides, and M. Khajavikhan, Science {\bf 346},  975 (2014).
\bibitem{r11}
 L. Feng, Z.J. Wong, R.-M. Ma, Y. Wang, and X. Zhang, Science {\bf 346}, 972 (2014).
 \bibitem{r12}
 M. Wimmer, A. Regensburger,	M.-A. Miri,	C. Bersch, D.N. Christodoulides, and U. Peschel, Nature Commun. {\bf 6}, 7782 (2015).
 \bibitem{r13}
 S.A.R. Horsley, M. Artoni, and G.C. La Rocca, Nature Photon. {\bf 9}, 436 (2015).
 \bibitem{r13b}
  S. Longhi, D. Gatti, and G. Della Valle, Sci. Rep. {\bf 5}, 13376  (2015).
 \bibitem{r14}
S. Longhi, Opt. Lett. {\bf 35}, 3844 (2010).
 \bibitem{r15}
M. Kulishov, J.M. Laniel, N. Belanger, J. Azana, and D.V. Plant, Opt. Express {\bf 13}, 3068 (2005).
 \bibitem{r16}
Z. Lin, H. Ramezani, T. Eichelkraut, T. Kottos, H. Cao, and D.N. Christodoulides, Phys. Rev. Lett. {\bf 106}, 213901 (2011). 
\bibitem{r17}
S. Longhi, J. Phys. A {\bf 44}, 485302 (2011).
\bibitem{r18}
H.F. Jones, J. Phys. A {\bf 45}, 1751 (2012).
\bibitem{r19}
L. Feng, Y.-L. Xu, W.S. Fegadolli, M.-H. Lu, J.E.B. Oliveira, V.R. Almeida, Y.-F. Chen, and A. Scherer, Nature Mat. {\bf 12}, 108 (2013).
\bibitem{r20}
M. Kulishov, H. F. Jones, and B. Kress, Opt. Express {\bf 23}, 9347 (2015). 
\bibitem{20bis}
A. Mostafazadeh, Phys. Rev. A {\bf 89}, 012709 (2014).
\bibitem{r21}
S. Longhi, Phys. Rev. A {\bf 82}, 032111 (2010).
\bibitem{r22}
S. Longhi and G. Della Valle, Ann. Phys. {\bf 334}, 35 (2013).
\bibitem{r23}
A. Ruschhaupt, F. Delgado, and J.G. Muga, J. Phys. A  {\bf 38}, L171 (2005).
\bibitem{r24}
R. El-Ganainy, K. G. Makris, D. N. Christodoulides, and Z.H. Musslimani, Opt. Lett. {\bf 32}, 2632 (2007).
\bibitem{r25}
S. Longhi, Phys. Rev. A {\bf 81}, 022102 (2010).
\bibitem{r26}
L. Poladian, Phys. Rev. E {\bf 54}, 2963 (1996).
\bibitem{r27}
S. Longhi, Phys. Rev. Lett. {\bf 105}, 013903 (2010).



\end{thebibliography}
\end{document}